\documentclass[aps,prl,twocolumn,groupedaddress,showpacs,superscriptaddress,nobibnotes]{revtex4}
\usepackage{graphicx}
\usepackage{epsfig}

\begin{document}

\title{Size Dependence and Convergence of the Retrieval Parameters of Metamaterials}

\author{Jiangfeng Zhou}
\affiliation{Ames Laboratory and Department of Physics and
Astronomy, Iowa State University, Ames, Iowa
50011}
\affiliation{Department of Electrical and Computer Engineering
and Microelectronics Research Center, Iowa State University, Ames,
Iowa 50011}

\author{Thomas Koschny}

\affiliation{Ames Laboratory and Department of Physics and
Astronomy, Iowa State University, Ames, Iowa 50011}
\affiliation{Institute of Electronic Structure and Laser -
Foundation for Research and Technology Hellas (FORTH),and Department
of Materials Science and Technology, University of Crete, Greece}

\author{Maria Kafesaki}
\affiliation{Institute of Electronic Structure and Laser -
Foundation for Research and Technology Hellas (FORTH),and Department
of Materials Science and Technology, University of Crete, Greece}

\author{Costas M. Soukoulis}
\affiliation{Ames Laboratory and Department of Physics and
Astronomy, Iowa State University, Ames, Iowa 50011}
\affiliation{Institute of Electronic Structure and Laser -
Foundation for Research and Technology Hellas (FORTH),and Department
of Materials Science and Technology, University of Crete, Greece}

\begin{abstract}
We study the dependence of the retrieval parameters, such as the
electric permittivity, $\epsilon$, the magnetic permeability, $\mu$,
and the index of refraction, $n$, on the size of the unit cell of a
periodic metamaterial. The convergence of the retrieved parameters
on the number of the unit cells is also examined. We have
concentrated our studies on the so-called fishnet structure, which
is the most promising design to obtain negative $n$ at optical
wavelengths. We find that as the size of the unit cell decreases,
the magnitude of the retrieved effective parameters increases. The
convergence of the effective parameters of the fishnet as the number
of the unit cells increases is demonstrated but found to be slower
than for regular split ring resonators and wires structures. This is
due to a much stronger coupling between the different unit cells in
the fishnet structure.
\end{abstract}


\maketitle


\section{Introduction}
The recent development of metamaterials with negative refractive
index has confirmed that structures can be fabricated that can be
interpreted as having both a negative effective permittivity,
$\epsilon$, and a negative effective permeability, $\mu$,
simultaneously. Since the original microwave experiments for the
demonstration of negative index behavior in the split ring
resonators (SRRs) and wires structures, new designs have been
introduced, such as the fishnet, that have pushed the existence of
the negative refractive index at optical wavelengths
\cite{Adv_Mat_soukoulis_review_2006,Sci_soukoulis_2007,Nat_Phonotics_shalaev_2007}.
Most of the experiments
\cite{PRL_zhang_2005_137404,dolling_OL_2006,science_soukoulis_2006,
Opl_Chettiar_2007,PRB_kafesaki_2007_235114} with the fishnet
structure measure the transmission, $T$, and the reflection, $R$,
and use the retrieval procedure
\cite{PRB_smith_retrieval_2002,PRE_koschny_retrieval_2003,
PRB_koschny_retrieval_2005,PRE_Smith_retrieval_2005,PRE_Chen_retrieval_2004}
to numerically obtain the effective parameters, $\epsilon$, $\mu$
and $n$. The only direct experimental way that was able to measure
both the phase and the group velocity, was done by pulse
measurements [6]. In all the other experiments only $T$ and/or $R$
are measured and then the retrieval procedure is used to obtain the
effective parameters. There is an unresolved issue in the retrieval
procedure, what is the size of the unit cell that needs to be used.
Most of the retrieval procedures use only one unit cell and made the
assumption that if more than one unit cell will be used, the results
will be the same as for the single unit cell. This is an untested
assumption and Koschny et al. [11] have systematically studied this
assumption for the SRRs and wires. It was found that this assumption
is correct if the wavelength is 30 times larger than the size of the
unit cell. No systematic study have been done on the dependence of
the retrieval parameters on the size of the unit cell of the fishnet
structure. In this work, we tackle this problem and also study the
convergence of the effective parameters from the retrieval procedure
as the number of the unit cells increases. We also, examine how the
figure of merit (FOM), which is defined as
$-\mathrm{Re}(n)/\mathrm{Im}(n)$, changes with the size of the unit
cell and the number of the unit cells.

\section{Numerical calculations}

\begin{figure}[htb]\centering
  \includegraphics[width=8cm]{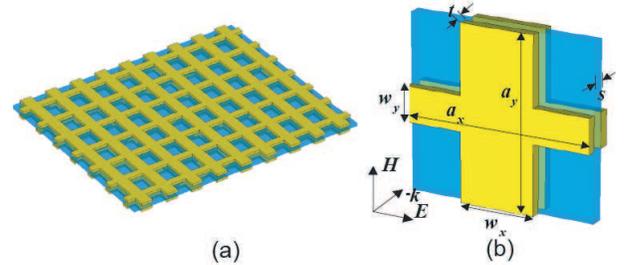}\\
  \caption{Schematic of a fishnet structure (a); a single
   unit cell with geometric parameters marked on it (b).
  \label{fig1_structure}}
\end{figure}

\begin{figure}[htb]\centering
  \includegraphics[width=6cm]{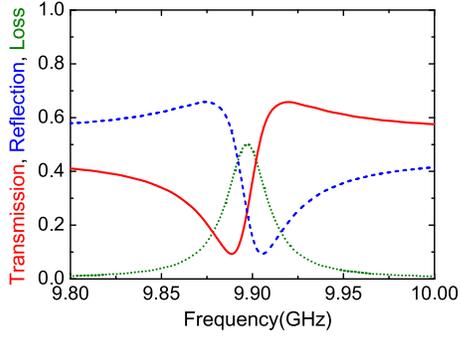}\\
  \caption{Simulated transmission (red solid), reflection (blue dashed)
  and loss (green dotted), which is proportional to absorption, of a fishnet structure.
  The geometric parameters are
  $a_x=a_y=15$ mm, $a_z=4$ mm, $w_x=6$ mm, $w_y=4$ mm, $s=0.5$ mm and $t=0.05$
  mm and the dielectric constant of the spacer is $\epsilon=5$.
  \label{fig2_T_R_A}}
\end{figure}

Our numerical simulations were done with CST Microwave Studio
(Computer Simulation Technology GmbH, Darmstadt, Germany), which
uses a finite-integration technique, and Comsol Multiphysics, which
uses a frequency domain finite element method. The schematics of the
unit cell of fishnet structure are shown in Fig.
\ref{fig1_structure}(b). Notice that he propagation direction is
perpendicular to the plane of the fishnet. In Fig. \ref{fig2_T_R_A}
we present the results for the transmission, $T$, reflection, $R$,
and absorption, $A=1-T-R$, for a given size of the unit cell.

\begin{figure}[htb]\centering
  \includegraphics[width=6cm]{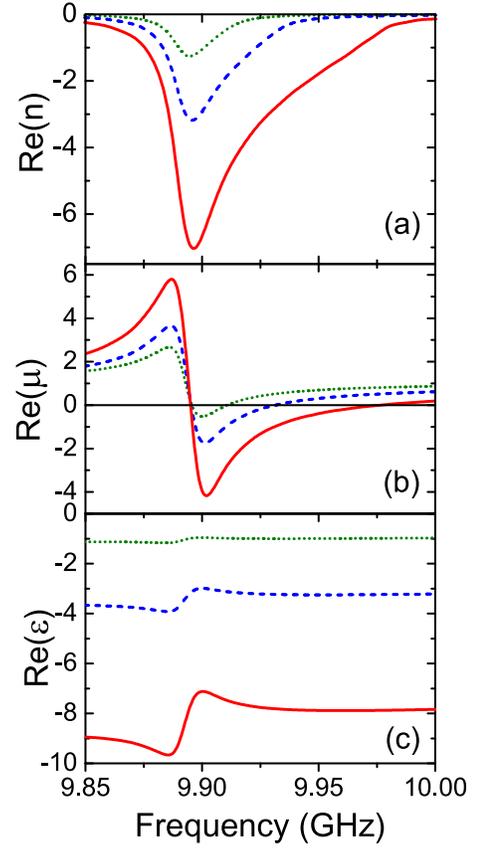}\\
  \caption{Retrieved real part of refractive index, $n$, (a), permeability,
  $\mu$, (b), and permittivity, $\epsilon$, (c),
 from simulated data shown in Fig. \ref{fig2_T_R_A}, using unit cell
   size in propagation direction $a_z=1$ mm (red solid),
  $a_z=2$ mm (blue dashed) and $a_z=4$ mm (green dotted). \label{fig3_n_mu}}
\end{figure}

In Fig. \ref{fig3_n_mu}, we present the results of the real part of
the retrieved effective parameters, $n$, $\mu$ and $\epsilon$, as a
function of the frequency. Notice that as the size of the unit cell
is getting larger the strength of the resonance of the magnetic
permeability $\mu$ is getting weaker. The same is true for the
magnitude of the index of refraction, $n$, which is also negative.
For the dimensions we have chosen, the resonance frequency is 9.9
GHz, which is equivalent to a wavelength of $\lambda=30.3$ mm. So
the ratio of $\lambda/a_z\approx$ 30, 15, 7.6 for the three sizes
($a_z=$1 mm, 2 mm and 4 mm) of the unit cell examined in Fig.
\ref{fig3_n_mu}. The real part of $\epsilon$ is always negative, and
shows an anti-resonance \cite{PRE_koschny_retrieval_2003}
corresponding to the magnetic resonance at 9.9 GHz.

\begin{figure}[htb]\centering
  \includegraphics[width=6cm]{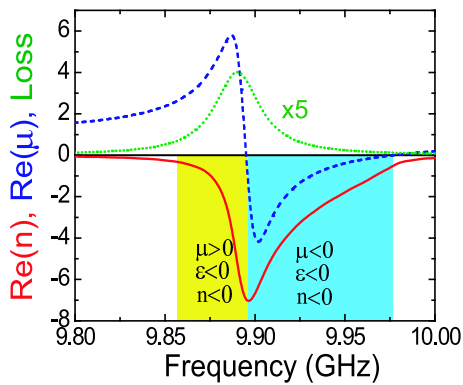}\\
  \caption{Retrieved real part of effective refractive index, Re($n$), (red solid),
  permeability, $\mathrm{Re}(\mu)$, (blue dashed),
  and normalized loss, $(1-T-R)/(1-R)$, (green
  dotted) (normalized loss has been magnified by a factor of 5 to improve
  visibility). The shadow regions, below 9.9 GHz (yellow) and above 9.9 GHz (blue),
  show frequency regions with $\mathrm{Re}(n)<-0.5$.
  \label{Fig_8_N_mu_loss}}
\end{figure}

Another important issue \cite{Nano_Photonic_penciu_2006} that the
community of metamaterials has not completely understood is why the
width of the negative $n$ is wider than the width of the negative
$\mu$, as can be seen in Fig. \ref{fig3_n_mu}(a) and
\ref{fig3_n_mu}(b). In Fig. \ref{Fig_8_N_mu_loss}, we plot together
the $\mu$ and $n$ (real parts) around the magnetic resonance
frequency of 9.9 GHz for the $a_z=1 mm$ case. Notice that $\mu<0$
for frequencies higher than 9.9 GHz. However $n$ is also negative
below 9.9 GHz where $\mu>0$. So we have two regions that give
negative $n$. The one above 9.9 GHz (drawn as blue in Fig.
\ref{Fig_8_N_mu_loss}) which has both $\epsilon$ and $\mu$ negative
and therefore $n$ is also negative. The other region, below 9.9 GHz
(shown as yellow in Fig. \ref{Fig_8_N_mu_loss}) has $\epsilon<0$ but
$\mu>0$ and still give $n<0$. The reason is that one can obtain
$\mathrm{Re}(n)$ negative provided that
$\mathrm{Im}(\epsilon)\mathrm{Im}(z)>\mathrm{Re}(\epsilon)\mathrm{Re}(z)$,
where $z=\sqrt{\mu/\epsilon}$. For the details of this explanation
see Ref.
\cite{PRB_zhou_cwp,Lakhtakia_n_sign_2002,Lakhtakia_n_sign_2004}. The
normalized loss, $(1-T-R)/(1-R)$, which is the ratio between the
absolute loss, $1-T-R$, and the total energy entering the fishnet
structure, $1-R$, is also shown in Fig. \ref{Fig_8_N_mu_loss}. The
normalized loss is a better measure of the absorption in a medium,
since it removes the effect of the reflection due to the impendence
mismatch at the interface between the medium and the vacuum space,
which results in lower absolute loss. If we compare the frequencies
with same $\mathrm{Re}(n)$ value in the regions below 9.9 GHz
(yellow) and above 9.9 GHz (blue), one can see that the normalized
loss in the region below 9.9 GHz (yellow) is much higher than the
region above 9.9 GHz (blue). The transmission, $T$, as shown in Fig.
\ref{fig2_T_R_A}, in region above 9.9 GHz (blue) is much higher than
in the region below 9.9 GHz (yellow). Therefore, the frequency
region above 9.9 GHz (blue) is the frequencies range which is
suitable for the applications of the negative refractive index
medium.


\begin{figure}[htb]\centering
  \includegraphics[width=6cm]{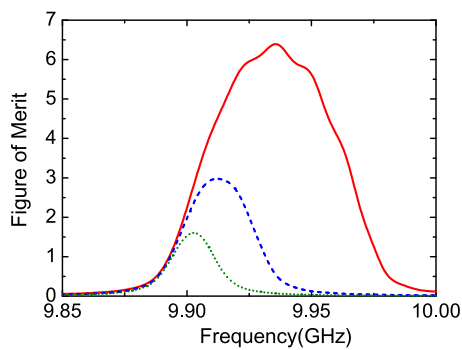}\\
  \caption{
 Figure of merit, $\mathrm{FOM}=-\mathrm{Re}(n)/\mathrm{Im}(n)$,
 for unit cell size $a_z=1$ mm (red solid),
  $a_z=2$ mm (blue dashed) and $a_z=4$ mm (green dotted).\label{fig4_FOM}}
\end{figure}

In Fig. \ref{fig4_FOM}, we present the results of the figure of
merit (FOM=$-\mathrm{Re}(n)/\mathrm{Im}(n)$) as the size of the unit
cell of the fishnet structure increases. Notice that as the size of
the unit cell is getting smaller, the FOM increases and reaches a
value of 6.5, when the size of the unit cell along the propagation
direction is equal to 1 mm. Notice also that the maximum of the
figure of the merit is higher than the resonance frequency, which is
9.9 GHz. The reason is that the $\mathrm{Im}(n)$ takes its larger
value at the resonance frequency and decreases as one goes away from
the resonance frequency. This is the reason that the FOM=6.5 for
$a_z=$ 1 mm has its maximum value at 9.95 GHz.

\begin{figure}[htb]\centering
  \includegraphics[width=6cm]{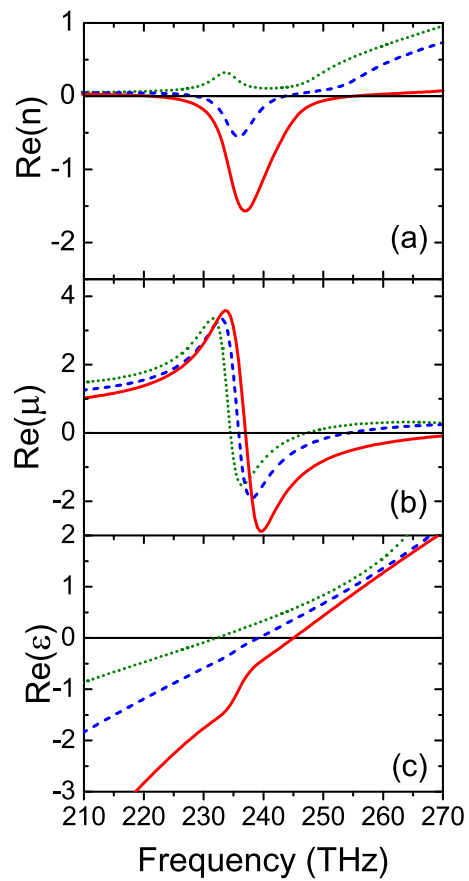}\\
  \caption{Retrieved real part of effective refractive index, $n$, (a), permeability,
  $\mu$, (b) and  permittivity, $\epsilon$, (c) from
  simulation data of a fishnet structure with parameters $a_z=200$ nm (red solid),
  $a_z=300$ nm (blue dashed), and $a_z=400$ nm (green dotted). The
  other geometry parameters are $a_x=a_y=600$ nm, $w_x=400$ nm,
  $w_y=200$ nm, $s=100$ nm and $t=60$ nm, and the dielectric
  constant of the spacer is $\epsilon=1.9$.
  \label{fig5_THz_n_eps_mu}}
\end{figure}

In Fig. \ref{fig5_THz_n_eps_mu}, we present the results of the
retrieved $n$, $\mu$ and $\epsilon$ (real parts only), as a function
of frequency, for THz frequencies. As we have discussed above, the
fishnet structure is the only structure that has been proven to give
negative $n$ at THz and optical frequencies
\cite{Sci_soukoulis_2007,Nat_Phonotics_shalaev_2007,Opl_Chettiar_2007,dolling_OL_2007_three_layer_LHM}.
To obtain negative $\mu$ and negative $n$ at THz frequencies, we
have to scale down the dimensions given in Fig. \ref{fig2_T_R_A}.
The size of the unit cell along the propagation direction $a_z=$ 200
nm, 300 nm and 400 nm. The resonance frequency is approximately 235
THz, which is equivalent to a wavelength $\lambda$ = 1.28 $\mu m$.
So the ratio of $\lambda/a_z=$6.4, 4.27 and 3.2 for the three sizes
$a_z=$200 nm, 300 nm, and 400 nm of the unit cell examined in Fig.
\ref{fig5_THz_n_eps_mu}. Notice in Fig. \ref{fig5_THz_n_eps_mu}(a)
that as the size of the unit cell increases the retrieved $n$ goes
from negative to positive values. The reason for the positive value
of $n$ is the following: As the size of the unit cell along the
propagation direction increases, the effective plasma frequency of
the fishnet structure, as shown in Fig. \ref{fig5_THz_n_eps_mu}(c),
decreases and becomes smaller than the magnetic resonance frequency
of $\mu$, as shown in Fig. \ref{fig5_THz_n_eps_mu}(b). Therefore, we
don't have an overlap of the negative region of $\epsilon$ and
$\mu$.

\begin{figure}[htb]\centering
  \includegraphics[width=6cm]{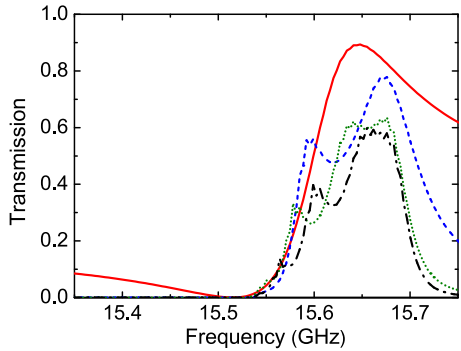}\\
  \caption{Simulated transmission spectra of one layer (red solid),
  four layers (blue dashed), eight layers (green dotted) and ten
  layers (black dash-dotted) of the fishnet structure. The geometric
  parameters are $a_x=a_y=10$ mm, $a_z=4$ mm, $w_x=4$ mm, $w_y=3$ mm,
  $s=1$ mm and $t=0.05$ mm, and the dielectric constant of the spacer
  is $\epsilon=4$.
  \label{fig6_T}}
\end{figure}

\begin{figure}[htb]\centering
  \includegraphics[width=6cm]{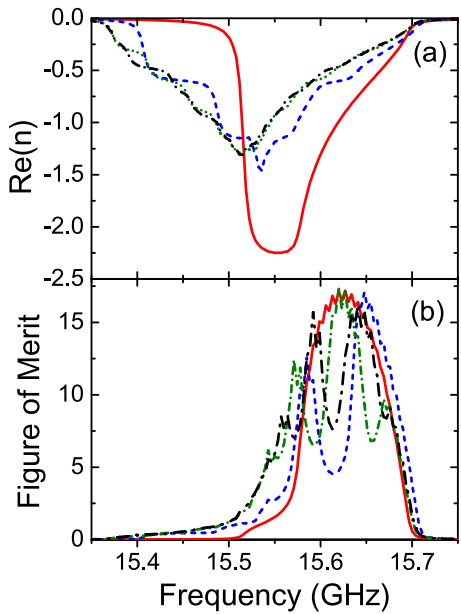}\\
  \caption{Retrieved real part of effective refractive index,$n$, and
  figure of merit of the simulated structure described in Fig. \ref{fig6_T}.
  \label{fig7_converge}}
\end{figure}

In Fig. \ref{fig6_T}, we present the results of the transmission of
the one unit cell, four unit cells, eight unit cells and ten unit
cells (along propagation direction) for a fishnet structure that has
a magnetic resonance at 15.65 GHz. Notice that as the number of the
unit cells increases, the transmission results converge to a common
value. This is seen more clearly in Fig \ref{fig7_converge}, where
we present the retrieved results for the effective index of
refraction. One can clearly see that the results for eight and ten
unit cells are exactly the same. This convergent length dependence
is a required property to attribute effective medium behavior to a
metamaterial \cite{PRB_koschny_retrieval_2005}. Notice that the
converged results for the effective parameters are different from
the effective parameters results of the one unit cell. As one can
see from Fig. \ref{fig7_converge}(a), the retrieved results for the
effective parameter $n$ for more than one unit cells are completely
different for the results of the one unit cell. This is true,
especially when the $\mathrm{Re}(n)$ is relative large. At the high
frequency edges of the negative $n$ region, there is no much
difference between the results of the one unit cell with the results
of more than one unit cells. In Fig. \ref{fig7_converge}(b), one
notices that the FOM does not change dramatically as one uses more
unit cells to do the retrieval procedure. As one uses more unit
cells, the FOM has more oscillations as a function of frequency.

\begin{figure}[htb]\centering
  \includegraphics[width=6cm]{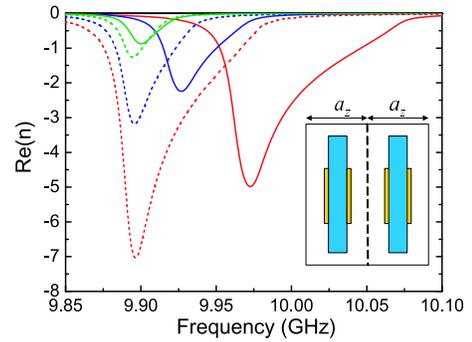}\\
  \caption{Retrieved real part of refractive index, $n$, from simulated data using unit cell
   size in propagation direction $a_z=1$ mm (red),
  $a_z=2$ mm (blue) and $a_z=4$ mm (green). Both one layer (dashed)
  and two layers (solid) results are shown.
  \label{Fig_9_N_1and2_layer}}
\end{figure}

Fig. \ref{Fig_9_N_1and2_layer} shows the real parts of effective
refractive index, $\mathrm{Re}(n)$, as a function of frequency for
one layer and two layers fishnet structure with different unit cell
size: $a_z=1$ mm, $a_z=2$ mm and $a_z=4$ mm. The resonance
frequencies, i.e. frequency with maximum $|\mathrm{Re}(n)|$, are
9.926 GHz (red dashed), 9.896 GHz (blue dashed) and  9.894 GHz
(green dashed) for one layer; 9.973 GHz (red solid), 9.927 GHz (blue
solid) and 9.900 GHz (green solid) for two layers. The difference in
value of resonance frequencies between one layer and two layer
simulations are 77 MHz (red, with $a_z=1$ mm), 31 MHz (blue, with
$a_z=2$ mm) and 6 MHz (green, with $a_z=4$ mm).
%
It is shown that the resonance frequency, $f_m$, of two layers
results decreases as $a_z$ increases, while $f_m$ of the one layer
only shifts slightly, when the size of the unit cell increases. The
frequency shift of two layers simulation is due to the coupling
between the two neighboring layers of the fishnet. Notice that as
the unit cell size $a_z $ increases, the coupling between neighboring
layers becomes weaker, and therefor the effective refractive index,
$\mathrm{Re}(n)$, of two layers approaches the one layer simulation
results.

\section{Conclusions}
We have systematically studied the size dependence of the retrieved
parameters, $\mu$, $\epsilon$ and $n$, of the fishnet metamaterial
structures as a function of the size of the unit cell. We found that
the retrieval parameters have a {\bf stronger} resonance behavior as
the size of the unit cell decreases. We have also studied the
convergence of the retrieved parameters as the number of the unit
cells increases. We found that the convergence depends on the ratio
of $\lambda/a$, where $\lambda$ is the wavelength and $a$ is the
size of the unit cell. The larger $\lambda/a$ is, the easier the
convergence. The convergence of the fishnet structure is much slower
than that observed in SRRs and/or wires systems. Finally, we gave an
explanation of why the frequency width where $n$ is negative is
wider than the width where $\mu$ is negative.

\section{Acknowledgments}%
Work at Ames Laboratory was supported by Dept. of Energy (Basic
Energy Sciences) under contract No. DE-AC02-07CH11358, by the AFOSR
under MURI grant (FA9550-06-1-0337), by Dept. of Navy, office of
Naval Research (Award No. N0014-07-1-0359), EU FET projects
Metamorphose and PHOREMOST, and by Greek Ministry of Education
Pythagoras project.




\end{document}